\documentclass[11pt]{article}
\usepackage{amsbsy,amsfonts}
\DeclareMathAlphabet{\bi}{OML}{cmm}{b}{i}

\newcommand{\bG}{\bi{G}}
\newcommand{\JPA}{\textit{J. Phys. A: Math. Gen.}}

\newcommand{\fl}{}
\newcommand{\etal}{\textit{et al.}}
\newcommand{\be}{\begin{equation}}
\newcommand{\ee}{\end{equation}}
\newcommand{\ba}{\begin{array}}
\newcommand{\ea}{\end{array}}
\newcommand{\p}{\partial}
\newcommand{\bu}{\bi{u}}

\newcommand{\G}{\bi{G}}

\def\ord{\mathop{\rm ord}\nolimits}

\def\ker{\mathop{\rm ker}\nolimits}
\def\Im{\mathop{\rm Im}\nolimits}
\def\diag{\mathop{\rm diag}\nolimits}
\newcommand{\bg}{\boldsymbol{\gamma}}
\newcommand{\bz}{\boldsymbol{\zeta}}
\def\Im{\mathop{\rm Im}\nolimits}

\def\diag{\mathop{\rm diag}\nolimits}

\def\wt{\mathop{\rm wt}\nolimits}
\newcommand{\ds}{\displaystyle}
\newtheorem{theo}{Theorem}
\newtheorem{lem}{Lemma}
\newtheorem{cor}{Corollary}
\newtheorem{prop}{Proposition}

\begin{document}

\title
{\protect\vspace*{-12mm}{\bf Why nonlocal recursion operators\\
produce local symmetries:\\ new results and
applications\protect\thanks{Published in {\em J. Phys. A: Math.
Gen.} {\bf 38} (2005) 3397--3407 \copyright 2005 IOP Publishing
Ltd\protect\\ \hspace*{4.5mm} Original article is available at {\tt
http://stacks.iop.org/JPhysA/38/3397}}}}

\author{A. Sergyeyev\\[3mm]
Mathematical Institute, Silesian University in Opava,\\ Na
Rybn\'\i{}\v{c}ku 1, 746\,01 Opava, Czech Republic\\ E-mail: {\tt
Artur.Sergyeyev@math.slu.cz}}
\date{}
\maketitle \protect\vspace*{-12mm}
\begin{abstract}
It is well known that integrable hierarchies in (1+1) dimensions are local
while the recursion operators that generate these hierarchies usually contain nonlocal terms.
We resolve this apparent discrepancy
by 
providing simple and universal sufficient conditions for a (nonlocal)
recursion operator in (1+1) dimensions to generate a hierarchy of local symmetries.
These conditions are satisfied by virtually all known today recursion operators
and are much easier to verify than those found in earlier work.
\looseness=-1

%

We also give explicit formulas for the nonlocal
parts of higher recursion, Hamiltonian and symplectic  operators
of integrable systems in (1+1) dimensions.
\looseness=-1

Using these two results we prove,
under some natural assumptions, the Maltsev--Novikov conjecture stating that
higher Hamiltonian, symplectic and recursion operators
of integrable systems in (1+1) dimensions are weakly nonlocal,
i.e., the coefficients of these operators are local
and these operators contain at most one integration operator in each term.
\looseness=-1

\bigskip



\end{abstract}

\section*{Introduction}

It is a common knowledge that
an integrable system of PDEs never comes alone --
it always is a member of an infinite integrable hierarchy.
In particular, if we deal with the evolution systems then the members of the hierarchy
are symmetries one for another, and using a recursion operator,
which maps symmetries to symmetries, offers a natural way to construct
the whole infinite hierarchy from a single seed system,
see e.g.\ \cite{olv_eng2, bl, dor} and references therein
and cf.\ \cite{bl, dor, ff2, oevth, ff3, sergromp}
and references therein
for the hierarchies generated by master symmetries.
\looseness=-1


The overwhelming majority of recursion operators in (1+1) dimensions
share two key features \cite{olv_eng2, bl, dor, wang}: they are {\em hereditary}, i.e., their
Nijenhuis torsion vanishes \cite{ff}, and {\em  weakly
nonlocal} \cite{mn}, i.e., all their nonlocal terms have the form $a
\otimes D^{-1} \circ b$, where $a$ and $b$ are local functions,
possibly vector-valued, and $D$ is the operator of total
$x$-derivative, see below for details.

On the other hand, it is well known that nearly all integrable hierarchies in (1+1)
dimensions are {\em local}. Usually it is not difficult to check
that applying the recursion operator to a local seed symmetry once
or twice yields local quantities, but the locality of the whole
{\em infinite} hierarchy is quite difficult to rigorously verify.
\looseness=-1

It is therefore natural to ask \cite{sw} whether a weakly nonlocal hereditary
operator will always produce a local hierarchy, as in the earlier
work \cite{dor}, \cite{sw}--\cite{serg2}
one always had to require the existence of some nontrivial additional structures
(e.g., the scaling symmetry \cite{sw, kras, serg2} or bihamiltonian structure \cite{dor, olv})
in order to get the proof of locality
through. \looseness=-1
We show that this is not necessary:
Theorem~\ref{th_loc1} below 
states that if for a normal weakly
nonlocal hereditary recursion operator $\mathfrak{R}$ the Lie
derivative $L_{\bi{Q}}(\mathfrak{R})$ of $\mathfrak{R}$ along a local symmetry $\bi{Q}$ vanishes
\footnote{
Where does the condition $L_{\bi{Q}}(\mathfrak{R})=0$ come from?
As all members of an integrable hierarchy must be compatible,
the symmetries $\mathfrak{R}^i(\bi{Q})$ must commute, and this is ensured by
requiring that $\mathfrak{R}$ be hereditary and that
$L_{\bi{Q}}(\mathfrak{R})=0$, cf.\  e.g.\ \cite{ff}. Moreover, $L_{\bi{Q}}(\mathfrak{R})=0$
means that $\mathfrak{R}$ is a recursion operator for the evolution system
$\bi{u}_{\tau}=\bi{Q}$.
\looseness=-1
}
and $\mathfrak{R}(\bi{Q})$ is local then
$\mathfrak{R}^j (\bi{Q})$ are local for all
$j=2,3,\dots$. Notice that, unlike e.g.\ \cite{sw, serg2},
we do {\em not} require the hierarchy in question
to be time-in\-de\-pen\-dent, and our Proposition~\ref{th_loc} and Theorem~\ref{th_loc1}
can be successfully employed for proving locality of
the so-called variable coefficients hierarchies, including for instance
those constructed in \cite{bl87, fu} and \cite{bl}, cf.\ Example 2 below.
\looseness=-1


Given an operator $\mathfrak{R}$, it is usually immediate
whether it is weakly nonlocal, but
it can be quite difficult to check whether it is hereditary, especially if we deal with
newly discovered integrable systems with no multi-Hamiltonian
representation and no Lax pair known. \looseness=-1
Amazingly enough, the existence of a scaling symmetry shared by
$\mathfrak{R}$ and $\bi{Q}$ enables us to avoid the cumbersome
direct verification of whether $\mathfrak{R}$ is hereditary and
allows to prove locality {\em and} commutativity of the
corresponding hierarchy in a very simple and straightforward manner,
as shown in Proposition~\ref{hercon1} and Corollary~\ref{hercon2}
below.
This is 
in a sense reminiscent of the construction of compatible Hamiltonian
operators via infinitesimal deformations in Smirnov \cite{smi} (see
also \cite{serg3} and references therein) and is quite different
from the approach of \cite{sw}, where {\em both}
$\mathfrak{R}$ being hereditary
and existence of scaling symmetry were required {\em ab initio}.
\looseness=-1

Let $\mathfrak{R}$, $\mathfrak{P}$, and $\mathfrak{S}$ be
respectively recursion, Hamiltonian and symplectic operator
for some
(1+1)-dimensional integrable system, and let all of
them be weakly nonlocal. 
Motivated by the examples of nonlinear Schr\"odinger and KdV
equations, Maltsev and Novikov \cite{mn} conjectured that higher
recursion operators $\mathfrak{R}^k$, higher Hamiltonian operators
$\mathfrak{P}\circ \mathfrak{R^\dagger}^k$ and higher symplectic
operators $\mathfrak{S}\circ \mathfrak{R}^k$ are weakly nonlocal for
all $k\in\mathbb{N}$ as well. \looseness=-1

Combining our Corollary~\ref{corloc} with the results of Enriquez,
Orlov and Rubtsov \cite{rubts} enabled us to prove this conjecture
under some natural assumptions, the most important of which is that
$\mathfrak{R}$ is hereditary, see Theorem~\ref{nmconj} below for
details. This has interesting and quite far-reaching consequences
for both theory and applications of integrable systems, e.g., in
connection with the so-called Whitham averaging, cf.\ discussion in
\cite{mn,m2,m3}. \looseness=-1

\section{Preliminaries}
Denote by $\mathcal{A}_{j}$ the algebra of locally analytic
functions of $x,t,\bu,\bu_{1},\dots,\bu_{j}$ under the standard
multiplication, and let
$\mathcal{A}=\bigcup_{j=0}^{\infty}\mathcal{A}_{j}$. We shall
refer to the elements of $\mathcal{A}$ as to {\em local}
functions \cite{mik1, sok, s, mik}. Here $\bu_{k}=(u_{k}^{1},\dots,u_{k}^{s})^{T}$ are
$s$-component vectors, $\bi{u}_0\equiv\bi{u}$, and the superscript $T$ stands
for the matrix transposition. The derivation \cite{olv_eng2, sok}
\looseness=-1
$$
D\equiv D_x=\frac{\p}{\p x}+\sum\limits_{j=0}^{\infty}\bu_{j+1}\cdot
\frac{\p}{\p\bu_j}
$$
makes $\mathcal{A}$ into a differential algebra. Informally, $x$ plays the role of
the space variable, and $D$ is the total $x$-derivative, cf.\ e.g.\
\cite{olv_eng2, sok}.
It is closely related to the operator of variational derivative
\cite{olv_eng2, bl, dor}
$$
\frac{\delta}{\delta\bu}=\sum\limits_{j=0}^{\infty}(-D)^{j}\frac{\p}{\p\bu_j}.
$$
In particular, see e.g.\ \cite{olv_eng2, dor},
for any $f\in\mathcal{A}$ we have
\be\label{vard}
\frac{\delta f}{\delta\bi{u}}=0 \quad \mbox{if and only if}\quad f\in\Im D.
\ee
Here and below `$\cdot$' stands for the scalar product of two
$s$-component vectors, and $\Im D$ denotes the image of $D$ in $\mathcal{A}$, so $f\in\Im D$
means that $f=D(g)$ for some $g\in\mathcal{A}$.
\looseness=-1

For a (scalar, vector or matrix) local function $f$ define
\cite{olv_eng2} its {\em order} $\ord f$ as the greatest integer $k$
such that $\p f/\p\bi{u}_k\neq 0$ (if $f=f(x,t)$, we set $\ord f=0$
by definition), and define the {\em directional derivative} of $f$
(cf.\ e.g.\ \cite{olv_eng2, ff}) by the formula
$$
f'=\sum\limits_{i=0}^{\infty}\frac{\p f}{\p\bu_{i}}D^{i}.
$$

Consider now the algebra $\mathrm{Mat}_{q}(\mathcal{A})[\![D^{-1}]\!]$
of formal series of the form $\mathfrak{H}=\sum
_{j=-\infty}^{k}h_{j}D^{j}$, where $h_{j}$ are $q\times q$ matrices
with entries from $\mathcal{A}$. The multiplication law in this
algebra is given by the (extended by linearity) generalized Leibniz
rule \cite{olv_eng2, mik1, s, mik}: \be\label{lei} a D^{i}\circ b
D^{j} =a \sum\limits_{q=0}^{\infty}{\displaystyle \frac{i(i-1)\cdots
(i-q+1)}{q!}}D^{q}(b)D^{i+j-q}. \ee The commutator $\left[
\mathfrak{A}, \mathfrak{B} \right]=\mathfrak{A} \circ \mathfrak{B}-
\mathfrak{B} \circ \mathfrak{A}$ further makes
$\mathrm{Mat}_{q}(\mathcal{A})[\![D^{-1}]\!]$ into a Lie algebra.

Recall \cite{olv_eng2, mik1, s, mik} that the {\em degree} $\deg\mathfrak{H}$ of
$\mathfrak{H}=\sum_{j=-\infty}^{p}h_{j}D^{j}\in\mathrm{Mat}_{q}(\mathcal{A})[\![D^{-1}]\!]$ is the
greatest integer $m$ such that $h_{m}\neq 0$.
For any $\mathfrak{H}=\sum_{j=-\infty}^{m}h_{j}D^{j}\in\mathrm{Mat}_{q}(\mathcal{A})[\![D^{-1}]\!]$
define its differential part $\mathfrak{H}_+=\sum_{j=0}^{m}h_{j}D^{j}$ and
nonlocal part $\mathfrak{H}_-=\sum_{j=-\infty}^{-1}h_{j}D^{j}$ so
that $\mathfrak{H}_-+\mathfrak{H}_+=\mathfrak{H}$,
and let $\mathfrak{H}^{\dagger}=\sum
_{j=-\infty}^{m}(-D)^{j}\circ
h_{j}^{T}$ stand for the formal adjoint of $\mathfrak{H}$,
see e.g.\ \cite{olv_eng2, mik1, s, mik}.
\looseness=-1
%

We shall employ the notation
$\mathcal{A}^q$ for the space of
$q$-component functions with entries from $\mathcal{A}$,
no matter whether they are interpreted as column or row vectors.
Following \cite{mn} we shall call
$\mathfrak{H}\in\mathrm{Mat}_{q}(\mathcal{A})[\![D^{-1}]\!]$
{\em weakly nonlocal} if there exist $\vec f_\alpha\in\mathcal{A}^q$,
$\vec g_\alpha\in\mathcal{A}^q$ and
$k\in\mathbb{N}$ such that
$\mathfrak{H}_{-}$ can be written in the form
$\mathfrak{H}_{-}=\sum_{\alpha=1}^{k}
\vec f_\alpha\otimes D^{-1}\circ \vec g_\alpha$.
We shall further say
that $\mathfrak{H}\in\mathrm{Mat}_{q}(\mathcal{A})[\![D^{-1}]\!]$
is {\em local} (or {\em purely differential}) if $\mathfrak{H}_{-}=0$.
Nearly all known today recursion operators in (1+1) dimensions,
as well as Hamiltonian and symplectic operators,
are weakly nonlocal, cf.\ e.g.\ \cite{wang}.
\looseness=-1

The space $\mathcal{V}$ of $s$-component columns with entries from $\mathcal{A}$
is made into a Lie algebra if we set
$[\bi{P},\bi{Q}]=\bi{Q}'(\bi{P})-\bi{P}'(\bi{Q})$, see e.g.\ \cite{olv_eng2, bl, ff, mik1}.
The Lie derivative of $\bi{R}\in\mathcal{V}$
along $\bi{Q}\in\mathcal{V}$ is then given \cite{olv_eng2, bl, dor, wangth} by
$L_{\bi{Q}}(\bi{R})=[\bi{Q},\bi{R}]$. The natural dual of $\mathcal{V}$
is the space $\mathcal{V}^*$ of $s$-component rows with entries from $\mathcal{A}$.
For $\boldsymbol{\gamma}\in\mathcal{V}^*$ we define \cite{bl, dor, sw, wangth} its Lie derivative
along $\bi{Q}\in\mathcal{V}$ as $L_{\bi{Q}}(\boldsymbol{\gamma})
=\boldsymbol{\gamma}'(\bi{Q})+\bi{Q}'^{\dagger}(\boldsymbol{\gamma})$, see
\cite{dor, wangth} for more details and for the related complex of formal
calculus of variations.
\looseness=-1

For $\bi{Q}\in\mathcal{V}$ and $\bg\in\mathcal{V}^*$ we have,
see e.g.\ \cite{olv_eng2},
$\delta(\bi{Q}\cdot\bg)/\delta\bi{u}=\bi{Q}'^\dagger(\bg)
+\bg'^\dagger(\bi{Q})$,
hence if $\bg'^\dagger(\bi{Q})=\bg'(\bi{Q})$ then
\be\label{lievar}
L_{\bi{Q}}(\bg)=\delta(\bi{Q}\cdot\bg)/\delta\bi{u}.
\ee
%




If $\mathfrak{R}:\mathcal{V}\rightarrow\mathcal{V}$,
$\mathfrak{S}:\mathcal{V}\rightarrow\mathcal{V}^*$,
$\mathfrak{P}:\mathcal{V}^*\rightarrow\mathcal{V}$,
$\mathfrak{N}:\mathcal{V}^*\rightarrow\mathcal{V}^*$
are weakly nonlocal or, even more broadly, belong to
$\mathrm{Mat}_{s}(\mathcal{A})[\![D^{-1}]\!]$, then we
can \cite{bl, oevth, ff} define their Lie derivatives along $\bi{Q}\in\mathcal{V}$
as follows: $L_{\bi{Q}}(\mathfrak{R})=\mathfrak{R}'[\bi{Q}]-[\bi{Q}',\mathfrak{R}]$,
$L_{\bi{Q}}(\mathfrak{N})=\mathfrak{N}'[\bi{Q}]+[\bi{Q}'^\dagger,\mathfrak{N}]$,
$L_{\bi{Q}}(\mathfrak{P})=\mathfrak{P}'[\bi{Q}]-\bi{Q}'\circ \mathfrak{P}-\mathfrak{P}\circ \bi{Q}'^\dagger$,
$L_{\bi{Q}}(\mathfrak{S})=\mathfrak{S}'[\bi{Q}]
+\bi{Q}'^\dagger \circ \mathfrak{S}+\mathfrak{S}\circ\bi{Q}'$, where
for
$\mathfrak{H}=\sum_{j=-\infty}^{m}h_{j}D^{j}$
we set $\mathfrak{H}'[\bi{Q}]=\sum_{j=-\infty}^{m}h'_{j}[\bi{Q}]D^{j}$.
Here and below we do {\em not} assume $\mathfrak{R}$ and
$\mathfrak{S}$ (resp.\ $\mathfrak{P}$ and $\mathfrak{N}$) to be
defined on the whole of $\mathcal{V}$ (resp.\ on the whole of
$\mathcal{V}^*$). \looseness=-1

An operator $\mathfrak{R}:\mathcal{V}\rightarrow\mathcal{V}$
is called {\em hereditary} \cite{ff} (or {\em Nijenhuis}
\cite{dor}) on a linear subspace $\mathcal{L}$
of the domain of definition of $\mathfrak{R}$ if for
all $\bi{Q}\in\mathcal{L}$
\be\label{hered}
L_{\mathfrak{R}(\bi{Q})}(\mathfrak{R})=\mathfrak{R}\circ
L_{\bi{Q}}(\mathfrak{R}).
\ee
In what follows
by saying that $\mathfrak{R}$ is hereditary without specifying $\mathcal{L}$
we shall mean that $\mathfrak{R}$
is hereditary on its whole domain of definition,
cf.\ e.g.\ \cite{ff}. If $\mathfrak{R}$ is
hereditary on $\mathcal{L}$, then
for any $\bi{Q}\in\mathcal{L}$ such that
$\mathfrak{R}^{k}(\bi{Q})\in\mathcal{L}$ for all $k\in\mathbb{N}$ we have
$[\mathfrak{R}^{i}(\bi{Q}),\mathfrak{R}^{j}(\bi{Q})]=0$,
$i,j=0,1,2,\dots$, cf.\ e.g.\ \cite{bl, oevth}.
We do not address here the issue of
proper definition of $\mathfrak{R}^j(\bi{Q})$ and refer
the reader to \cite{guthrie, marv, swro}
and \cite{as-sb} and references therein for details.
\looseness=-1

Denote by $\mathcal{S}(\mathfrak{R},\bi{Q})$ the linear span of
$\mathfrak{R}^i(\bi{Q})$, $i=0,1,2,\dots$. We readily see from~(\ref{hered})
that $L_{\mathfrak{R}^i(\bi{Q})}(\mathfrak{R})=0$ for all
$i=0,1,2,\dots$
if and only if $L_{\bi{Q}}(\mathfrak{R})=0$ and
$\mathfrak{R}$ is hereditary on
$\mathcal{S}(\mathfrak{R},\bi{Q})$. Hence,
if
$L_{\mathfrak{R}^i(\bi{Q})}(\mathfrak{R})=0$ for
all $i=0,1,2,\dots$ then $[\mathfrak{R}^{i}(\bi{Q}),\mathfrak{R}^{j}(\bi{Q})]=0$ for all
$i,j=0,1,2,\dots$.

\section{The main result and its applications}

Consider a weakly nonlocal operator
$\mathfrak{R}:\mathcal{V}\rightarrow\mathcal{V}$ of the form
\be\label{ro}
\mathfrak{R}=\sum\limits_{i=0}^{r}a_{i}D^{i}+\sum
\limits_{\alpha=1}^{p}\G_{\alpha}\otimes D^{-1}\circ \bg_{\alpha},
\ee
where $a_i$ are $s\times s$ matrices with entries from
$\mathcal{A}$,
$\bG_{\alpha}\in\mathcal{V}$, $\bg_{\alpha}\in\mathcal{V}^*$,
and $r\geq 0$.



We shall call $\mathfrak{R}$ of the form (\ref{ro}) {\em normal\/}
if for all $\alpha,\beta=1,\dots,p$ we have
$\bg_\alpha'=\bg_\alpha'^\dagger$,
$\boldsymbol{\zeta}_\alpha'=\boldsymbol{\zeta}_\alpha'^\dagger$,
where $\boldsymbol{\zeta}_\alpha=\mathfrak{R}^\dagger(\bg_\alpha)$,
and $L_{\bG_\alpha}(\bg_\beta)=0$. This is a very common property:
it appears that all known today weakly nonlocal hereditary recursion
operators of integrable systems in (1+1) dimensions are normal.

\begin{prop}\label{th_loc}
Consider a normal $\mathfrak{R}:\mathcal{V}\rightarrow\mathcal{V}$
of the form (\ref{ro}), and let
$\bi{Q}\in\mathcal{V}$  and $\mathfrak{R}$ be such that $\mathfrak{R}$ is hereditary
on $\mathcal{S}(\mathfrak{R},\bi{Q})$,
$L_{\bi{Q}}(\mathfrak{R})=0$,
and $L_{\bi{Q}}(\bg_{\alpha})=0$ for all $\alpha=1,\dots,p$.

Then $\bi{Q}_j=\mathfrak{R}^j(\bi{Q})$ are local and commute for all
$j=0,1,2,\dots$.
\end{prop}
\noindent {\em Proof.} The commutativity of $\bi{Q}_j$ immediately
follows from $\mathfrak{R}$ being hereditary on $\mathcal{S}(\mathfrak{R},\bi{Q})$,
see above. Now assume that $\bi{Q}_j$ is local and $L_{\bi{Q}_j}(\bg_\alpha)=0$,
and let us show that $\bi{Q}_{j+1}$ is local
and $L_{\bi{Q}_{j+1}}(\bg_\alpha)=0$.
First of all, by (\ref{lievar}) we have
$\delta(\bi{Q}_j\cdot\bg_\alpha)/\delta\bi u=L_{\bi{Q}_j}(\bg_\alpha)=0$,
so by (\ref{vard})
$\bi{Q}_j\cdot\bg_\alpha\in\Im D$ for all $\alpha=1,\dots,p$,
and hence $\bi{Q}_{j+1}=\mathfrak{R}(\bi{Q}_j)$ is local.
\looseness=-1

To proceed, we need the following lemma:
\begin{lem}\label{simple}
Let $\mathfrak{R}:\mathcal{V}\rightarrow\mathcal{V}$
of the form (\ref{ro}) and $\bi{Q}\in\mathcal{V}$
be such that $L_{\bG_\alpha}(\bg_\beta)=0$, $L_{\bi{Q}}(\bg_\alpha)=0$,
and $\bg_\alpha'^\dagger(\bi{Q})=\bg_\alpha'(\bi{Q})$ for all $\alpha,\beta=1,\dots,p$.

Then $L_{\mathfrak{R}(\bi{Q})}(\bg_\alpha)=
\delta(\bi{Q}\cdot\mathfrak{R}^\dagger(\bg_\alpha))/\delta\bi{u}$
for all $\alpha=1,\dots,p$.
\end{lem}
{\em Proof of the lemma.}
As $\bg_\alpha'^\dagger(\bi{Q})=\bg_\alpha'(\bi{Q})$, by (\ref{lievar})
we have
$\delta(\bi{Q}\cdot\bg_\alpha)/\delta\bi{u}
=L_{\bi{Q}}(\bg_\alpha)=0$,
so 
by (\ref{vard})
$\bi{Q}\cdot\bg_\alpha=D(f_\alpha)$
for some $f_\alpha\in\mathcal{A}$.
Likewise, $\bG_\beta\cdot\bg_\alpha=D(g_{\alpha\beta})$
for some $g_{\alpha\beta}\in\mathcal{A}$, whence
\looseness=-1
$$
\mathfrak{R}(\bi{Q})\cdot\bg_\alpha=\bi{Q}\cdot\mathfrak{R}^\dagger(\bg_{\alpha})+
D\left(\sum_{i=1}^{r}\sum_{j=0}^{i-1}(-D)^{j}(a_{i}^{T}\bg_{\alpha})\cdot
D^{i-j-1}(\bi{Q})+\sum_{\beta=1}^{p}g_{\alpha\beta}f_{\beta}\right).
$$
Using this formula along with (\ref{vard}) and (\ref{lievar}) yields
$L_{\mathfrak{R}(\bi{Q})}(\bg_\alpha)=
\delta(\mathfrak{R}(\bi{Q})\cdot\bg_\alpha)/\delta\bi{u}=
\delta(\bi{Q}\cdot\mathfrak{R}^\dagger(\bg_\alpha))/\delta\bi{u}$.
The lemma is proved. $\square$

As $\mathfrak{R}$ is hereditary on $\mathcal{S}(\mathfrak{R},\bi{Q})$,
repeatedly using (\ref{hered})
yields $L_{\bi{Q}_j}(\mathfrak{R})=L_{\mathfrak{R}^j(\bi{Q})}(\mathfrak{R})=
\mathfrak{R}^j \circ L_{\bi{Q}}(\mathfrak{R})=0$.
Next, using Lemma~\ref{simple}, the normality of $\mathfrak{R}$, the equality
$\boldsymbol{\zeta}_\alpha'=\boldsymbol{\zeta}_\alpha'^\dagger$,
where $\boldsymbol{\zeta}_\alpha=\mathfrak{R}^\dagger(\bg_\alpha)$,
and (\ref{lievar}), we obtain
$L_{\bi{Q}_{j+1}}(\bg_\alpha)=L_{\mathfrak{R}(\bi{Q}_j)}({\bg_\alpha})=
L_{\bi{Q}_j}(\mathfrak{R}^\dagger(\bg_\alpha))=L_{\bi{Q}_j}(\mathfrak{R}^\dagger)
\bg_\alpha+\mathfrak{R}^\dagger L_{\bi{Q}_j}(\bg_\alpha)=
L_{\bi{Q}_j}(\mathfrak{R}^\dagger)
\bg_\alpha=(L_{\bi{Q}_j}(\mathfrak{R}))^\dagger\bg_\alpha=0$.
The induction on $j$
starting from $j=0$ completes the proof.
$\square$
\looseness=-1

If $\boldsymbol{G}_\alpha$, $\alpha=1,\dots,p$,
are linearly independent over the field $\mathbb{T}$ of locally analytic functions of $t$
(notice that this can always be assumed without loss of generality),
then the conditions $L_{\bi{Q}}(\bg_{\alpha})=0$, $\alpha=1,\dots,p$,
are equivalent to the requirement that $\mathfrak{R}(\bi{Q})$ is local, and we arrive at
the result announced in Introduction.

\begin{theo}\label{th_loc1}
Let $\boldsymbol{G}_\alpha$, $\alpha=1,\dots,p$, be linearly
independent over the field $\mathbb{T}$ of locally analytic
functions of $t$. Suppose that a normal weakly nonlocal
$\mathfrak{R}: \mathcal{V}\rightarrow \mathcal{V}$ of the form
(\ref{ro}) and $\bi{Q}\in\mathcal{V}$ are such that
$L_{\bi{Q}}(\mathfrak{R})=0$, $\mathfrak{R}$ is hereditary on
$\mathcal{S}(\mathfrak{R},\bi{Q})$, and $\mathfrak{R}(\bi{Q})$ is
local.

Then the quantities
$\bi{Q}_j=\mathfrak{R}^j(\bi{Q})$ are local for all $j=2,3,\dots$,
and $[\bi{Q}_j,\bi{Q}_k]=0$ for all $j,k=0,1,2\dots$.
\looseness=-1
\end{theo}
\noindent{\em Proof.}
By virtue of Proposition~\ref{th_loc} it is
enough to show that if $\mathfrak{R}(\bi{Q})$ is local then
$L_{\bi{Q}}(\bg_{\alpha})=0$ for all $\alpha=1,\dots,p$.
To prove this, suppose that $\mathfrak{R}(\bi{Q})$ is local but for some value(s)
of $\alpha$ we have $L_{\bi{Q}}(\bg_\alpha)\neq 0$.

Then we have $\mathfrak{R}(\bi{Q})=\bi{M}+\sum_{\alpha=1}^p \bi{G}_\alpha\omega_\alpha$, where $\bi{M}$ is local,
and $\omega_\alpha$ denotes the nonlocal part of $D^{-1}(\bg_\alpha\cdot\bi{Q})$ (some of $\omega_\alpha$ may be zeros).
By assumption, $\mathfrak{R}(\bi{Q})$ is local, so $\sum_{\alpha=1}^p \bi{G}_\alpha\omega_\alpha=0$.
Moreover, as $D^i(\mathfrak{R}(\bi{Q}))$, $i=1,2,\dots$, are local too, we arrive at the following system
of algebraic equations for $\omega_\alpha$:
\[
\sum_{\alpha=1}^p D^j(\bi{G}_\alpha)\omega_\alpha=0,\quad j=0,1,2,\dots.
\]
This system has the same structure as (\ref{lindep}), and using the linear independence of $\bi{G}_\alpha$
over $\mathbb{T}$ we conclude, in analogy with the proof of Lemma~\ref{lem1} from Appendix, that
$\omega_\alpha=0$ for all $\alpha=1,\dots,p$. Hence $\bg_\alpha\cdot\bi{Q}\in\Im D$ and by (\ref{vard}) we have
$\delta(\bg_\alpha\cdot\bi{Q})/\delta\bi{u}=0$. Finally, as $\bg_\alpha'^\dagger=\bg'_\alpha$ by assumption,
(\ref{lievar}) yields
$L_{\bi{Q}}(\bg_{\alpha})=0$ for all $\alpha=1,\dots,p$,
as required. \looseness=-1
$\square$

The seed symmetry $\bi{Q}$ often commutes with $\bi{G}_\alpha$:
$L_{\bi{Q}}(\bi{G}_\alpha)\equiv [\bi{Q},\bi{G}_\alpha]=0$. Then we can bypass the check
of the conditions $L_{\bi{Q}}(\bg_{\alpha})=0$ in Proposition~\ref{th_loc} as follows.
\looseness=-1
\begin{cor}\label{propg}
If $\boldsymbol{G}_\alpha$, $\alpha=1,\dots,p$,
are linearly independent over the field $\mathbb{T}$ of locally analytic functions of $t$,
then for any $\mathfrak{R}$ of the form (\ref{ro}) and any $\bi{Q}\in\mathcal{V}$ such that
$L_{\bi{Q}}(\mathfrak{R})=0$ and $L_{\bi{Q}}(\G_{\alpha})=0$ for all $\alpha=1,\dots,p$ we have
$L_{\bi{Q}}(\bg_{\alpha})=0$, $\alpha=1,\dots,p$.
\looseness=-1
\end{cor}
\noindent{\em Proof.}
Indeed,
$(L_{\bi{Q}}(\mathfrak{R}))_-=\sum_{\alpha=1}^{p}(\G_{\alpha}\otimes
D^{-1}\circ L_{\bi{Q}}(\boldsymbol{\gamma}_{\alpha})
+L_{\bi{Q}}(\G_{\alpha})\otimes D^{-1}\circ
\boldsymbol{\gamma}_{\alpha}) =\sum_{\alpha=1}^{p}\G_{\alpha}\otimes
D^{-1}\circ L_{\bi{Q}}(\boldsymbol{\gamma}_{\alpha})$. As
$L_{\bi{Q}}(\mathfrak{R})=0$ implies
$(L_{\bi{Q}}(\mathfrak{R}))_-=0$, we get
$\sum_{\alpha=1}^{p}\G_{\alpha}\otimes D^{-1}\circ
L_{\bi{Q}}(\boldsymbol{\gamma}_{\alpha})=0$,
whence by linear
independence of $\G_\alpha$ over $\mathbb{T}$ and Lemma~\ref{lem1}
(see Appendix) we obtain $L_{\bi{Q}}(\boldsymbol{\gamma}_{\alpha})=0$, as
required. \looseness=-1
$\square$

We also have the following `dual' of Proposition~\ref{th_loc} for the elements of $\mathcal{V}^*$.
\begin{prop}\label{th_loc_cosym}
Consider a hereditary operator
$\mathfrak{R}:\mathcal{V}\rightarrow\mathcal{V}$
of the form (\ref{ro})
and assume that
$L_{\bG_\alpha}(\mathfrak{R})=0$
for all $\alpha=1,\dots,p$.
Let
$\bz\in\mathcal{V}^*$ be such that
$L_{\bG_{\alpha}}(\bz)=0$
for all $\alpha=1,\dots,p$, $\bz'=\bz'^\dagger$,
and $(\mathfrak{R}^\dagger(\bz))'=(\mathfrak{R}^\dagger(\bz))'^\dagger$.
\looseness=-1

Then $\bz_{j}=\mathfrak{R}^{\dagger j}(\bz)$  are local, i.e.,
$\bz_{j}\in\mathcal{V}^*$,
and satisfy $\bz'_{j}=\bz_{j}'^\dagger$
for all $j\in\mathbb{N}$.
\end{prop}

{\em Proof.}
Again, assume that $\bz_j$ is local,
$L_{\bG_{\alpha}}(\bz_{j})=0$,
and $\bz'_j=\bz_j'^\dagger$, and
let us prove that $\bz_{j+1}$ is local as
well, $L_{\bG_{\alpha}}(\bz_{j+1})=0$,
and $\bz'_{j+1}=\bz_{j+1}'^\dagger$.

As $\mathfrak{R}$ is hereditary,
the equalities $\bz'=\bz'^\dagger$ and
$(\mathfrak{R}^\dagger(\bz))'=(\mathfrak{R}^\dagger(\bz))'^\dagger$
imply \cite{oevth, ff} that
$\bz'_{j}=\bz_{j}'^\dagger$ for all $j=0,1,2,\dots$.
%
We further have
$L_{\bG_{\alpha}}(\bz_{j+1})=L_{\bG_{\alpha}}(\mathfrak{R}^\dagger\bz_{j})=
L_{\bG_{\alpha}}(\mathfrak{R}^\dagger)\bz_{j} +\mathfrak{R}^\dagger
L_{\bG_{\alpha}}(\bz_{j})=
L_{\bG_{\alpha}}(\mathfrak{R}^\dagger)\bz_{j}=
(L_{\bG_{\alpha}}(\mathfrak{R}))^\dagger\bz_{j}=0$, as desired.

Finally,
$\delta(\bz_j\cdot\bi{G}_{\alpha})/\delta\bi{u}=L_{\bG_{\alpha}}(\bz_{j})=0$
implies, by virtue of (\ref{vard}), that 
$\bi{G}_{\alpha}\cdot\bz_j\in\Im D$, and hence
$\bz_{j+1}$ is indeed local.
The induction on $j$ completes the proof. $\square$
\looseness=-1

\begin{cor}\label{corloc}
Let an operator
$\mathfrak{R}:\mathcal{V}\rightarrow\mathcal{V}$
of the form (\ref{ro})
be hereditary and normal, and let $L_{\bG_\alpha}(\mathfrak{R})=0$, $\alpha=1,\dots,p$.
\looseness=-1

Then $\bz_{\alpha,j}=\mathfrak{R}^{\dagger j}(\bg_\alpha)$ and
$\G_{\alpha,j}=\mathfrak{R}^j(\G_\alpha)$ are local,
$\bz'_{\alpha,j}=\bz_{\alpha,j}'^\dagger$, and $[\G_{\alpha,j}, \G_{\alpha,k}]=0$
for all $j,k=0,1,2,\dots$ and $\alpha=1,\dots,p$.
\looseness=-1
\end{cor}

\section{Hereditary operators and scaling}
Given an $\bi{S}\in\mathcal{V}$,
if $L_{\bi{S}}(K)=\kappa K$ for some constant $\kappa$, then $K$ is said to be of
weight $\kappa$ (with respect to the scaling $\bi{S}$), and we write $\kappa=\wt_{\bi{S}}(K)$, cf.\ e.g.\ \cite{sw}.

\begin{prop}\label{hercon1}Let
$\mathfrak{R}:\mathcal{V}\rightarrow\mathcal{V}$ and
$\bi{Q}\in\mathcal{V}$ be such that $L_{\bi{Q}}(\mathfrak{R})=0$.
Suppose that $\mathfrak{R}$ has the form (\ref{ro}),
$r\equiv\deg\mathfrak{R}>0$,
$L_{\bi{Q}}(\bg_\alpha)=0$ for all $\alpha=1,\dots,p$,
$q\equiv\ord\bi{Q}>\max(\ord a_r-r,1)$,
the matrix $\p\bi{Q}/\p\bi{u}_q$ has $s$ distinct eigenvalues,
and $\det\p\bi{Q}/\p\bi{u}_q\neq 0$.
Further assume that there exists a nonzero constant $\zeta$
and an $s$-component vector function $\bi{S}_0(\bi{u})$
such that for $\bi{S}=x\bi{u}_1+\bi{S}_0(\bi{u})$ we have
$L_{\bi{S}}(\mathfrak{R})=r\zeta\mathfrak{R}$,
$L_{\bi{S}}(\bi{Q})=q\zeta\bi{Q}$, and there exists an $s\times s$
matrix $\Gamma$ with entries from $\mathcal{A}$ that
simultaneously diagonalizes $\p\bi{Q}/\p\bi{u}_q$
and $\p\bi{S}_0/\p\bi{u}$ and satisfies
$\Gamma'[\bi{S}]-xD(\Gamma)=0$.

Then $L_{\mathfrak{R}^j(\bi{Q})}(\mathfrak{R})=0$  for all $j=1,2\dots$,
and hence $\mathfrak{R}$ is hereditary on $\mathcal{S}(\mathfrak{R},\bi{Q})$
and $[\mathfrak{R}^{i}(\bi{Q}),\mathfrak{R}^{j}(\bi{Q})]=0$ for all
$i,j=0,1,2,\dots$.
\end{prop}
{\em Proof.}
Consider an algebra $\tilde\mathcal{A}$ of all locally analytic functions that depend on
$x,t$, a finite number of $\bi{u}_j$, and a finite number
of nonlocal variables from
the universal Abelian covering over the system $\bi{u}_{\tau}=\bi{Q}$,
see \cite{as-sb, v2} and references therein for more details on this covering.
Let $\mathfrak{L}\equiv\sum_{i=-\infty}^m b_i D^i$,
where $b_i$ are $s\times s$ matrices with entries from
$\tilde\mathcal{A}$, satisfy $\mathfrak{L}'[\bi{Q}]-[\bi{Q}',\mathfrak{L}]=0$.
\looseness=-1

Assume first that $s=1$. Then, as $q>1$, equating to zero the
coefficient at $D^{m+q-1}$ in
$\mathfrak{L}'[\bi{Q}]-[\bi{Q}',\mathfrak{L}]=0$ yields $q
\p\bi{Q}/\p\bi{u}_q D(b_m) -m b_m D(\p\bi{Q}/\p\bi{u}_q)=0$, or
equivalently $D(b_m(\p\bi{Q}/\p\bi{u}_q)^{-m/q})=0$. In complete
analogy with Proposition~5 of \cite{as-sb}, the kernel of $D$ in
$\tilde\mathcal{A}$ is readily seen to be exhausted by the functions
of $t$ and $\tau$. Hence
$b_m=c_m(t,\tau)(\p\bi{Q}/\p\bi{u}_q)^{m/q}$ for some function
$c_m(t,\tau)$. \looseness=-1

For $s>1$ a similar computation shows that there exists (cf.\ e.g.\ \cite{mik1, s})
a diagonal $s\times s$ matrix $c_m(t,\tau)$
such that $b_m=\Gamma^{-1}c_m(t,\tau)\Lambda^{m/q}\Gamma$, where
$\Gamma$ is a matrix bringing $\p\bi{Q}/\p\bi{u}_q$ into the diagonal form,
i.e., $\Gamma\p\bi{Q}/\p\bi{u}_q\Gamma^{-1}=\diag(\lambda_1,\dots,\lambda_s)\equiv\Lambda$, where $\lambda_i$
are the eigenvalues of $\p\bi{Q}/\p\bi{u}_q$,
and $\Lambda^{m/q}=\diag(\lambda_1^{m/q},\dots,\lambda_s^{m/q})$.

It is straightforward to verify that
$\mathfrak{L}_j\equiv L_{\mathfrak{R}^j(\bi{Q})}(\mathfrak{R})$ for $j=1,2,\dots$
satisfy $L_{\bi{Q}}(\mathfrak{L}_j)\equiv\mathfrak{L}_j'[\bi{Q}]-[\bi{Q}',\mathfrak{L}_j]=0$.
Moreover, under the assumptions made $\mathfrak{R}$ is a recursion operator for
the system $\bi{u}_{\tau}=\bi{Q}$, and, as $L_{\bi{Q}}(\bg_\alpha)=0$ for all $\alpha=1,\dots,p$,
by Proposition 2 of \cite{as-sb} we have $\mathfrak{R}^j(\bi{Q})\in\tilde\mathcal{A}^s$ for all $j\in\mathbb{N}$.
Then, using the above formulas for the leading coefficients of $\mathfrak{L}_j$
and the condition $\Gamma'[\bi{S}]-xD(\Gamma)=0$ along with the assumption that $\Gamma$ diagonalizes
$\p\bi{S}_0/\p\bi{u}$, we readily find that $\wt_{\bi{S}}(\mathfrak{L}_j)=\zeta\deg\mathfrak{L}_j$.
\looseness=-1

As $q>1$, equating to
zero the coefficient at $D^{r+q}$ on the l.h.s.\ of $L_{\bi{Q}}(\mathfrak{R})=0$,
we conclude that the leading coefficient $\Phi\equiv\p\bi{Q}/\p\bi{u}_q$ of the formal series $\bi{Q}'$
commutes with the leading coefficient $a_r$ of $\mathfrak{R}$. Moreover, as $q>\ord a_r-r$,
the same is true for the leading coefficient $a_r^j\Phi$ of $(\mathfrak{R}^j(\bi{Q}))'$
for all $j=1,2,\dots$.
Therefore, the coefficient at $D^{jr+q}$ in $\mathfrak{L}_j$
vanishes, and $\deg(\mathfrak{L}_j)<q+rj$.
%
On the other hand, 
it is immediate that
$L_{\bi{S}}(\mathfrak{L}_j)
=(rj+q)\zeta \mathfrak{L}_j$.
This is in contradiction with
the formula
$\wt_{\bi{S}}(\mathfrak{L}_j)=\zeta\deg\mathfrak{L}_j$
unless $\mathfrak{L}_j=0$,
and the result follows. $\square$
\looseness=-1

{\em Remark.} 
The above proof can be readily extended
to include scalings $\bi{S}$ of more general form and
to handle the case when the coefficients of $\mathfrak{R}$
involve nonlocal variables from the universal Abelian covering over $\bi{u}_\tau=\bi{Q}$.
\looseness=-1

Theorem~\ref{th_loc1}
together with Propositions~\ref{th_loc} and \ref{hercon1}
yields the following assertion.
\looseness=-1

\begin{cor}\label{hercon2}Under the assumptions of Proposition~\ref{hercon1}
suppose that $\mathfrak{R}$ is normal,
and at least one of the following conditions is satisfied:
\looseness=-2
\begin{itemize}
\item[i)] $L_{\bi{Q}}(\bg_{\alpha})=0$, $\alpha=1,\dots,p$;
\item[ii)]$\boldsymbol{G}_\alpha$, $\alpha=1,\dots,p$,
are linearly independent over 
$\mathbb{T}$ and $L_{\bi{Q}}(\G_{\alpha})=0$, $\alpha=1,\dots,p$;
\item[iii)]$\boldsymbol{G}_\alpha$, $\alpha=1,\dots,p$,
are linearly independent over 
$\mathbb{T}$
and $\mathfrak{R}(\bi{Q})$ is local.
\end{itemize}
\looseness=-2

Then $\bi{Q}_j=\mathfrak{R}^j(\bi{Q})$ are local and commute
for all $j=0,1,2,\dots$.
\end{cor}

\section{Higher recursion, Hamiltonian and symplectic operators}

Consider an operator $\mathfrak{R}$ of the form (\ref{ro}) and another operator of similar form:
\be\label{ro1}
\tilde{\mathfrak{R}}=\sum\limits_{i=0}^{\tilde r}{\tilde a}_{i}D^{i}+
\sum\limits_{\alpha=1}^{\tilde p}
{\tilde\G}_{\alpha}\otimes
D^{-1}\circ
\tilde{\bg}_{\alpha}.
\ee
For a moment we do {\em not} assume that $\mathfrak{R}$ and $\tilde\mathfrak{R}$ act on $\mathcal{V}$, so
we do not specify whether the quantities $\G_\alpha$, $\bg_\alpha$, $\tilde{\G}_\alpha$, $\tilde{\bg}_\alpha$
belong to $\mathcal{V}$ or to $\mathcal{V}^*$.

Using the lemma from Section 2 of \cite{rubts}
we readily find that
\be
\label{prod}
(\mathfrak{R}\circ\tilde\mathfrak{R})_-=
\sum
\limits_{\alpha=1}^{\tilde p}\mathfrak{R}(\tilde\G_{\alpha})
\otimes D^{-1}\circ \tilde{\bg}_{\alpha}
+\sum
\limits_{\alpha=1}^{p}\G_{\alpha}\otimes D^{-1}
\circ \tilde\mathfrak{R}^{\dagger}(\bg_{\alpha}).
\ee

Repeatedly using
(\ref{prod})
yields the following formulas that hold for integer $n,m\geq 1$:
\begin{eqnarray}
\fl
(\mathfrak{R}^{n})_-=
\sum\limits_{j=0}^{n-1}{\ds\frac{(n-1)!}
{(n-1-j)!j!}}
\left(\sum
\limits_{\alpha=1}^{p}\mathfrak{R}^{j}(\G_{\alpha})
\otimes D^{-1}\circ (\mathfrak{R}^{\dagger})^{n-1-j}
(\bg_{\alpha})\right),\label{prod2}\\
\fl((\mathfrak{R}^{\dagger})^{n})_-=
-\sum\limits_{j=0}^{n-1}{\ds\frac{(n-1)!}
{(n-1-j)!j!}}
\left(\sum
\limits_{\alpha=1}^{p}\mathfrak{R}^{\dagger j}(\bg_{\alpha})
\otimes D^{-1}\circ \mathfrak{R}^{n-1-j}
(\G_{\alpha})\right),\label{prod2a}\\
\fl(\mathfrak{R}^{n}\circ\tilde\mathfrak{R}^{m})_-=
\sum\limits_{j=0}^{n-1}{\ds\frac{(n-1)!}
{(n-1-j)!j!}}
\left(\sum
\limits_{\alpha=1}^{p}\mathfrak{R}^{j}(\G_{\alpha})
\otimes D^{-1}\circ \tilde\mathfrak{R}^{\dagger m}
(\mathfrak{R}^{\dagger})^{n-1-j}(\bg_{\alpha})\right)\nonumber\\
+\sum\limits_{j=0}^{m-1}{\ds\frac{(m-1)!}
{(m-1-j)!j!}}
\left(\sum
\limits_{\alpha=1}^{\tilde p}
\mathfrak{R}^{n}\tilde\mathfrak{R}^{j}(\tilde\G_{\alpha})
\otimes D^{-1}\circ
(\tilde\mathfrak{R}^{\dagger})^{m-1-j}(\tilde{\bg}_{\alpha})
\right).\label{prod2b}
\end{eqnarray}

Corollary~\ref{corloc}, combined with
(\ref{prod})--(\ref{prod2b}), immediately yields the following result.
\begin{cor}\label{cor_hgh_op2}
Suppose that
$\mathfrak{R}:\mathcal{V}\rightarrow\mathcal{V}$
meets the requirements of Corollary~\ref{corloc}, and
$\mathfrak{P}~:\nobreak\mathcal{V}^*\nobreak\rightarrow\nobreak\mathcal{V}$,
$\mathfrak{S}:\mathcal{V}\rightarrow\mathcal{V}^*$,
$\mathfrak{N}:\mathcal{V}\rightarrow\mathcal{V}$,
$\mathfrak{T}:\mathcal{V}^*\rightarrow\mathcal{V}^*$
are purely differential operators.

Then $\mathfrak{R}^{k}$, $\mathfrak{R}^{\dagger k}$,
$\mathfrak{P}\circ \mathfrak{R}^{\dagger k}$, $\mathfrak{R}^{\dagger
k}\circ\mathfrak{S}$, $\mathfrak{S}\circ\mathfrak{R}^{k}$,
$\mathfrak{N}^q\circ \mathfrak{R}^{k}$, and
$\mathfrak{T}^q\circ\mathfrak{R}^{\dagger k}$ are weakly nonlocal
for all $k,q=0,1,2,\dots$.
\end{cor}

If $\mathfrak{B}$ is a scalar differential operator of degree $b$,
then \cite{sok}
$\dim_{\mathbb{T}}(\mathcal{A}\bigcap\ker\mathfrak{B})\leq b$, and
using Lemma~\ref{lem1} (see Appendix) we can readily prove the
following assertion.
\begin{cor}\label{cor_hgh_op3}
Let $s=1$. Assume that $\mathfrak{R}$ and
$\mathfrak{P}$ (resp.\ $\mathfrak{S}$) meet the requirements
of Corollary~\ref{cor_hgh_op2}, $\deg\mathfrak{P}=b$
(resp.\ $\deg\mathfrak{S}=b$),
and $\mathfrak{R}^{\dagger j}(\bg_{\alpha})$ (resp.\
$\mathfrak{R}^{j}(\G_{\alpha})$) are linearly
independent over $\mathbb{T}$ for all $j=0,\dots,n-1$ and $\alpha=1,\dots,p$.

Then there exist at most $[b/p]$ {\em local} linear combinations of
$\mathfrak{R}^{k}\circ\mathfrak{P}$
(resp.\ $\mathfrak{R}^{\dagger k}\circ\mathfrak{S}$), $k=1,\dots,n$, and
any such local linear combination involves only $\mathfrak{R}^{k}\circ\mathfrak{P}$
(resp.\ $\mathfrak{R}^{\dagger k}\circ\mathfrak{S}$)
with $k\leq [b/p]$.
\looseness=-1
\end{cor}

If $\mathfrak{P}$ is a Hamiltonian operator (resp.\ if $\mathfrak{S}$ is
a symplectic operator), the above results, especially Corollary~\ref{cor_hgh_op3},
enable us to obtain an estimate for the number of local, i.e., purely differential,
Hamiltonian (resp.\ symplectic) operators among the linear combinations of
$\mathfrak{R}^{k}\circ\mathfrak{P}$
(resp.\ $\mathfrak{R}^{\dagger k}\circ\mathfrak{S}$).
Such estimates play an important role
e.g.\ in the construction of Miura-type transformations
\cite{bl}.
\looseness=-1

Finally, using Propositions~\ref{th_loc} and \ref{th_loc_cosym}
we can readily generalize Corollary~\ref{cor_hgh_op2}
to the case of weakly nonlocal $\mathfrak{P}$,
$\mathfrak{S}$, $\mathfrak{T}$, $\mathfrak{N}$
as follows:
\begin{theo}\label{nmconj}
Suppose that
$\mathfrak{R}:\mathcal{V}\rightarrow\mathcal{V}$
of the form (\ref{ro})
meets the requirements of Corollary~\ref{corloc}, and
$\bi{K}_\beta,\bi{H}_\beta\in\mathcal{V}$ and
$\boldsymbol{\eta}_\beta,\boldsymbol{\zeta}_\beta\in\mathcal{V}^*$
are such that $L_{\bi{K}_\beta}(\mathfrak{R})=0$, $L_{\bi{H}_\beta}(\mathfrak{R})=0$,
$\boldsymbol{\eta}_\beta'=\boldsymbol{\eta}_\beta'^\dagger$,
$\boldsymbol{\zeta}_\beta'=\boldsymbol{\zeta}_\beta'^\dagger$,
$(\mathfrak{R}^\dagger(\boldsymbol{\eta}_\beta))'
=(\mathfrak{R}^\dagger(\boldsymbol{\eta}_\beta))'^\dagger$,
$(\mathfrak{R}^\dagger(\boldsymbol{\zeta}_\beta))'
=(\mathfrak{R}^\dagger(\boldsymbol{\zeta}_\beta))'^\dagger$,
$L_{\bi{K}_\beta}(\bg_\alpha)=0$, $L_{\bi{H}_\beta}(\bg_\alpha)=0$,
$L_{\bi{G}_\alpha}(\boldsymbol{\eta}_\beta)=0$, and $L_{\bi{G}_\alpha}(\boldsymbol{\zeta}_\beta)=0$
for all $\alpha=1,\dots,p$ and $\beta=1,\dots,m$. Further assume that
$\mathfrak{P}:\mathcal{V}^*\rightarrow\mathcal{V}$,
$\mathfrak{S}:\mathcal{V}\rightarrow\mathcal{V}^*$,
$\mathfrak{T}:\mathcal{V}^*\rightarrow\mathcal{V}^*$ and
$\mathfrak{N}:\mathcal{V}\rightarrow\mathcal{V}$ are weakly nonlocal and we have
$\mathfrak{P}_-=\sum_{\beta=1}^m \bi{K}_\beta\otimes D^{-1}
\circ \bi{H}_\beta$,
$\mathfrak{S}_-=\sum_{\beta=1}^m \boldsymbol{\zeta}_\beta\otimes
D^{-1}\circ \boldsymbol{\eta}_\beta$,
$\mathfrak{T}_-=\sum_{\beta=1}^m \boldsymbol{\zeta}_\beta\otimes D^{-1}
\circ \bi{K}_\beta$, and
$\mathfrak{N}_-=\sum_{\beta=1}^m \bi{H}_\beta\otimes
D^{-1}\circ \boldsymbol{\eta}_\beta$.

Then 
$\mathfrak{P}\circ\mathfrak{R}^{\dagger k}$,
$\mathfrak{T}\circ\mathfrak{R}^{\dagger k}$,
$\mathfrak{S}\circ\mathfrak{R}^{k}$,
and $\mathfrak{N}\circ\mathfrak{R}^{k}$
are weakly nonlocal
for all $k=0,1,2,\dots$.
\end{theo}

%
%
Notice that if $\mathfrak{P}$ is a Hamiltonian operator and $\mathfrak{S}$ is
a symplectic operator, then
they are skew-symmetric ($\mathfrak{P}^\dagger=-\mathfrak{P}$
and $\mathfrak{S}^\dagger=-\mathfrak{S}$),
and we can set without loss of generality
$\bi{H}_\beta=\epsilon_\beta \bi{K}_\beta$ and
$\boldsymbol{\zeta}_\beta=\tilde\epsilon_\beta \boldsymbol{\eta}_\beta$, where
$\epsilon_\beta$ and $\tilde \epsilon_\beta$
are constants taking one of three values, $-1$, $0$ or  $+1$,
see e.g.\ \cite{m3}.
The conditions of Theorem~\ref{nmconj}
for $\boldsymbol{\zeta}_\beta$ and $\bi{H}_\beta$ are then automatically satisfied.
Moreover, if $\mathfrak{R}$ is a recursion operator, $\mathfrak{P}$ a Hamiltonian operator,
and $\mathfrak{S}$ a symplectic
operator for an integrable system in (1+1) dimensions, then
Theorem~\ref{nmconj} proves, under some natural assumptions that are
satisfied for virtually all known examples, the Maltsev--Novikov
conjecture which states \cite{mn} that higher recursion operators
$\mathfrak{R}^{k}$, higher Hamiltonian operators
$\mathfrak{R}^{k}\circ\mathfrak{P}$, and higher symplectic operators
$\mathfrak{S}\circ\mathfrak{R}^{k}$ are weakly nonlocal for all
$k=0,1,2,\dots$. \looseness=-1

\section{Examples}
Consider a hereditary recursion operator (see e.g.\ the discussion at p.~122 of
\cite{wangth} and references therein)
$$
\mathfrak{R}=D^2+2 a u_1^2+\frac43 b u_1+c -\frac23(3 a u_1+b)
D^{-1}\circ u_2
$$
for the generalized potential modified Korteweg--de Vries equation
$$u_t=u_3+ au_1^3+b u_1^2 + c u_1,$$
where $a,b,c$ are arbitrary constants.
This operator meets the requirements of Theorem~\ref{th_loc} for
$\bi{Q}=u_1$, so all $\bi{Q}_j=\mathfrak{R}^j(\bi{Q})$,
$j=1,2,\dots$, are local.

The equation in question has infinitely many Hamiltonian operators
$\mathfrak{P}=D$ and
$\mathfrak{P}_j=\mathfrak{R}^j \circ \mathfrak{P}$, $j\in\mathbb{N}$ (in particular,
we have $\mathfrak{P}_1=D^3+(2 a u_1^2+\frac43 b u_1+c)D -\frac23 (3 a
u_1+b)u_2 \allowbreak +\frac23 (3 a u_1+b) D^{-1}\circ u_1$).
By Corollary~\ref{cor_hgh_op2}
all $\mathfrak{P}_j$, $j=1,2,\dots$, are weakly nonlocal, and by
Corollary~\ref{cor_hgh_op3} $\mathfrak{P}$ is the only {\em local}
Hamiltonian operator among $\mathfrak{R}^j \circ \mathfrak{P}$ for
$j=0,1,2\dots$.

For another example, consider a linear combination of the Harry Dym
equation and the time-independent parts of its scaling symmetries, cf.\ e.g.\ \cite{bl, fu, bl87}:
\be\label{nhd}
u_t=u^3 u_3 + a x u_1 + b u,
\ee
where $a$ and $b$ are arbitrary constants, and a hereditary
recursion operator for (\ref{nhd})
$$
\ba{l} \mathfrak{R}=\exp(-3(a + b) t)u^3 D^3\circ u \circ D^{-1}
\circ
\exp((a + b) t)/u^2\\
=\exp(- 2 (a + b) t)(u^2 D^2 - u u_1 D + u u_2) + \exp(-3(a + b)
t) u^3 u_3 D^{-1} \circ \exp((a + b) t)/u^2. \ea
$$
Again, the requirements of Theorem \ref{th_loc} are met for
$\bi{Q}=\exp(-3 (a+b)t)u^3 u_3$, so all
$\bi{Q}_j=\mathfrak{R}^j(\bi{Q})$, $j=1,2,\dots$, are local.

Notice that in both of these examples there is no scaling symmetry
of the form used in \cite{sw},
and hence the locality of corresponding hierarchies cannot
be established by direct application of the results from \cite{sw}.

\section*{Acknowledgements}
I am sincerely grateful to Profs. M. B\l aszak and V.V. Sokolov and
Drs. M. Marvan, M.V. Pavlov and R.G. Smirnov for stimulating
discussions.

This research was supported in part by the Jacob Blaustein Postdoctoral
Fellowship, the Czech Grant Agency (GA\v CR) under grant No.\ 201/04/0538,
by the Ministry of Education, Youth and Sports of Czech Republic
under grant MSM:J10/98:192400002 and under the development project
254/b for the year 2004, and by Silesian University in Opava
under the internal grant IGS 1/2004.

\looseness=-1


\appendix
\section*{Appendix}
\setcounter{section}{1}

Here we prove the following lemma
kindly communicated to the author by V V Sokolov.
\begin{lem}\label{lem1}
Consider
$\mathfrak{H}=\sum_{\alpha=1}^{m}
\vec f_{\alpha}\otimes D^{-1}\circ\vec g_\alpha$,
where $\vec f_{\alpha},\vec g_\alpha\in\mathcal{A}^q$, and
$\vec f_\alpha$ are linearly independent over the field
$\mathbb{T}$ of locally analytic functions of $t$.

Then $\mathfrak{H}=0$
if and only if $\vec g_{\alpha}=0$ for all $\alpha=1,\dots,m$.
\end{lem}
{\em Proof.}
Clearly, $\mathfrak{H}=0$ if and only if
$\mathfrak{H}^{\dagger}=0$.
%
Using (\ref{lei}) we find that
\[
\mathfrak{H}^{\dagger}=-\sum_{j=0}^{\infty}\sum_{\alpha=1}^{m}
(-1)^j\vec g_\alpha\otimes D^{j}(\vec f_{\alpha})D^{-1-j}.
\]
Equating to zero the coefficients at powers of $D$ in
$\mathfrak{H}^{\dagger}=0$, we obtain the following system of
linear {\em algebraic} equations for
$\vec g_\alpha$:
\be\label{sys}
\sum_{\alpha=1}^{m}g_\alpha^k D^{j}(f_{\alpha}^d)=0, \quad
d,k=1,\dots,q;j=0,1,2,\dots.
\ee

We want to prove that the linear independence of $\vec f_{\alpha}$
over $\mathbb{T}$ implies that
$g_\alpha^k=0$ for all $\alpha$ and $k$.
To this end let us first fix $k$ and consider (\ref{sys}) as a system
of linear equations for the components $g_\alpha^k$ of $\vec g_\alpha$.

Clearly, if the rank $\rho$ of the matrix of this system equals
$m$, then $g_\alpha^k=0$, so we can prove our claim by
proving that if $\rho < m$, then $\vec f_{\alpha}$ are {\em
linearly dependent} over $\mathbb{T}$.
Indeed, if $\rho<m$, then the columns of our matrix are
linearly dependent over $\mathcal{A}$.
On the other hand, $\rho$ of them must be
linearly independent over $\mathcal{A}$.
Assume without loss of generality that
these are just the first $\rho$ columns.
The rest can be expressed via them, that is,
there exist $h^\alpha_\beta\in\mathcal{A}$
such that
\be\label{lindep}
D^j(\vec f_\beta)=\sum\limits_{\alpha=1}^\rho h^\alpha_\beta D^j(\vec f_\alpha),
\beta=\rho+1,\dots,m, j=0,1,2,\dots.
\ee
As $h^\alpha_\beta$ are independent of $j$,
the consistency of the above equations and the linear independence of
first $\rho$ columns over $\mathcal{A}$ imply that
$D(h^\alpha_\beta)=0$, hence $h^\alpha_\beta=h^\alpha_\beta(t)$, and
(\ref{lindep}) for $j=0$ implies the linear dependence
of $\vec f_\alpha$ over $\mathbb{T}$,
which contradicts our initial assumptions.
\looseness=-1
Thus, if $\vec f_\alpha$, $\alpha=1,\dots,m$, are linearly independent
over $\mathbb{T}$, then the matrices in question are of rank $m$ for all $k$,
and hence $\vec g_\alpha=0$ for all $\alpha=1,\dots,m$.
$\square$



\begin{thebibliography}{99}
\footnotesize
\bibitem{olv_eng2}
Olver P J 1993 {\it Applications of Lie Groups to Differential
Equations} (New York: Springer)


\bibitem{bl}
B\l aszak M 1998 {\it Multi-Hamiltonian Theory of Dynamical
Systems} (Heidelberg: Springer)

\bibitem{dor}Dorfman I 1993 {\it Dirac Structures
and Integrability of Nonlinear Evolution Equations}
(Chichester: John Wiley \& Sons)

\bibitem{ff2} Fokas A S and Fuchssteiner B 1981
The hierarchy of the Benjamin-Ono equation
{\it Phys. Lett.} A {\bf 86} 341--5

\bibitem{oevth}
Oevel W 1984 {\it Rekursionmechanismen f\"ur Symmetrien und
Erhaltungss\"atze in Integrablen Systemen}, PhD thesis
(Paderborn: University of Paderborn) \looseness=-1

\bibitem{ff3} Finkel F and Fokas A S 2002
On the construction of evolution equations admitting a master symmetry
{\it Phys. Lett.} A {\bf 293}
36--44 ({\it Preprint} nlin.SI/0112002)

\bibitem{sergromp}Sergyeyev A 2002
On sufficient conditions of locality for hierarchies of symmetries of evolution systems
{\it Rep. Math. Phys.} {\bf 50} 307--314


\bibitem{wang} Wang J P 2002
A list of $1+1$ dimensional integrable equations and their properties
{\it J. Nonlinear Math. Phys.} {\bf 9}, suppl.~1, 213--33
\bibitem{ff}
Fuchssteiner B,  Fokas A S 1981
Symplectic structures, their B\"acklund
transformations and hereditary symmetries
{\em Physica D} {\bf 4}
47--66


\bibitem{mn}Maltsev A Ya and Novikov S P 2001
On the local systems Hamiltonian in
the weakly non-local Poisson brackets
{\it Physica D} {\bf 156}
53--80 ({\it Preprint} nlin.SI/0006030)

\bibitem{sw}
Sanders J A and Wang J P 2001
Integrable Systems and their Recursion Operators
{\em Nonlinear Analysis} {\bf 47}
5213--40

\bibitem{olv}Olver P J 1987
Bi-Hamiltonian systems, in:
{\it Ordinary and
partial differential equations (Dundee, 1986)}
(Harlow: Longman) pp~176--93

\bibitem{adl}Adler V E 1991
Lie-algebraic approach to nonlocal symmetries of integrable systems
{\em Theor.\ Math.\ Phys.} {\bf 89} 1239--48

\bibitem{serg1}Sergyeyev A 2004
Locality of symmetries generated by nonhereditary, inhomogeneous,
and time-dependent recursion operators:
a new application for formal symmetries
{\it Acta Appl. Math.} {\bf 83}
95--109 ({\it Preprint} nlin.SI/0303033)


\bibitem{kras}Krasil'shchik I S 2002
A simple method to prove locality of symmetry hierarchies
{\it Preprint} DIPS 9/2002 {\it available at} {\tt
http://www.diffiety.org}

\bibitem{serg2}Sergyeyev A 2004
The structure of cosymmetries
and a simple proof  of locality for hierarchies of
Symmetries for odd order evolution systems, in
{\it Symmetry in Nonlinear Mathematical Physics} (Kyiv: Institute of Mathematics of NASU)
Part 1, pp~238--245 {\it available online at}
{\tt
http://www.slac.stanford.edu/econf/C0306234/papers/sergyeyev.pdf}

\bibitem{bl87} B\l aszak M 1987
Soliton point particles of extended evolution equations
{\it J. Phys. A: Math. Gen.} {\bf 20} L1253--5

\bibitem{fu} Fuchssteiner B 1993
Integrable nonlinear evolution equations with time-dependent coefficients
{\it J. Math. Phys.} {\bf 34} 5140--58

\bibitem{smi}Smirnov R G 1997
Bi-Hamiltonian formalism: a constructive approach
{\it Lett. Math. Phys.} {\bf 41} 333--47

\bibitem{serg3}Sergyeyev A 2004
A simple way of making a Hamiltonian system into a bi-Hamiltonian one
{\it Acta Appl. Math.} {\bf 83} 183--97 ({\it Preprint}
nlin.SI/0310012)

\bibitem{rubts}Enriquez B, Orlov A  and Rubtsov V 1993
Higher Hamiltonian structures (the ${\rm sl}\sb 2$ case)
{\it JETP Lett.} {\bf 58}
658--64
({\it Preprint} hep-th/9309038)

\bibitem{m2} Maltsev A Ya 2005 Weakly-nonlocal symplectic structures,
Whitham method, and weakly-nonlocal symplectic structures of
Hydrodynamic Type \JPA {\bf 38} 637--82
({\it Preprint} nlin.SI/0405060)

\bibitem{m3} Maltsev A Ya 2002
The averaging of nonlocal
Hamiltonian structures in Whitham's method
{\it Int. J. Math. and Math. Sci.} 30 (2002), no. 7, 399--434 ({\it
Preprint} solv-int/9910011)

\bibitem{mik1}
Mikhailov A V, Shabat A B and Yamilov R I 1987
The symmetry approach to classification of
nonlinear equations. Complete lists of integrable systems
{\em Russ.\ Math.\ Surv.\/} {\bf 42}(4) 1--63
\bibitem{sok}Sokolov V V  1988 {\em Russ.\ Math.\ Surv.\/}
{\bf 43}(5) 165--204
\bibitem{s}
Mikhailov A V, Shabat A B and Sokolov V V 1991
The symmetry approach to classification of integrable equations,
in {\em What is Integrability?}, ed V E~Zakharov, (New York:
Springer) pp~115--84\looseness=-1
\bibitem{mik}
Mikhailov A V, Yamilov R I 1998
Towards classification of (2+1)-dimensional~integrable equations.
Integrability conditions. I
{\em J.\ Phys.\ A: Math.\ Gen.} {\bf 31} 6707--15


\bibitem{wangth} Wang J P 1998 {\it Symmetries and Conservation Laws of
Evolution Equations} PhD Thesis
(Amsterdam: Vrije Universiteit van Amsterdam)

\bibitem{guthrie}Guthrie G A 1994
Recursion operators and non-local symmetries
{\it Proc. Roy. Soc. London Ser. A} {\bf 446} (1926) 107--14
\bibitem{marv}Marvan M 1996
Another look on recursion operators
{\it Differential Geometry and Applications (Brno, 1995)} eds J
Jany\v ska \etal
(Brno: Masaryk University)
pp~393--402 {\it available at} {\tt
http://www.emis.de/proceedings}

\bibitem{swro}Sanders J A and Wang J P 2001
On Recursion Operators {\em Physica D} {\bf 149} 1--10

\bibitem{as-sb}
Sergyeyev A 2000  On recursion operators
and nonlocal symmetries of evolution equations
{\em Proc. Sem. Diff. Geom.}, ed D Krupka, Silesian
University in Opava, pp~159--73 ({\it Preprint} nlin.SI/0012011)

\bibitem{v2} Bocharov A V, Chetverikov V N, Duzhin S V, Khor'kova N G,
Krasil'shchik I S, Samokhin~A~V, Torkhov Yu N,
Verbovetsky A M and Vinogradov A M 1999 {\it Symmetries and Conservation Laws
for Differential Equations of Mathematical Physics}
(Providence, RI: American Mathematical Society)
\looseness=-2






\end{thebibliography}
\end{document}